\documentclass[journal,twoside,web]{ieeecolor}
\usepackage{tmi}
\usepackage{cite}
\usepackage{amsmath,amssymb,amsfonts}
\usepackage{algorithmic}
\usepackage{graphicx}
\usepackage{textcomp}
\let\labelindent\relax
\usepackage{enumitem}
\usepackage{url}

\usepackage{booktabs}
\usepackage{multirow}
\usepackage{multicol}
\usepackage{adjustbox}
\usepackage[table]{xcolor}
\usepackage[flushleft]{threeparttable}

\usepackage[linesnumbered,ruled,vlined]{algorithm2e}
\usepackage{amsmath}

\def\BibTeX{{\rm B\kern-.05em{\sc i\kern-.025em b}\kern-.08em
    T\kern-.1667em\lower.7ex\hbox{E}\kern-.125emX}}
\begin{document}
\title{IRS: Incremental Relationship-guided Segmentation for Digital Pathology}
\author{Ruining Deng, Junchao Zhu, Juming Xiong, Can Cui, Tianyuan Yao, Junlin Guo, Siqi Lu, Marilyn Lionts, Mengmeng Yin, Yu Wang, Shilin Zhao, Yucheng Tang, Yihe Yang, Paul Dennis Simonson, Mert R. Sabuncu, Haichun Yang, Yuankai Huo
\thanks{*Y. Huo is the contact author, e-mail: yuankai.huo@vanderbilt.edu}
\thanks{R. Deng, J. Zhu, J. Xiong, C. Cui, T. Yao, J. Guo, S. Lu, M. Lionts, and Y. Huo were with Vanderbilt University, Nashville, TN, 37215, USA}
\thanks{M. Yin, Y. Wang, S. Zhao, and H. Yang were with Vanderbilt University Medical Center, Nashville, TN, 37215, USA}
\thanks{Y. Tang was with NVIDIA Corp., Seattle, WA, 98105, USA}
\thanks{R. Deng, Y. Yang, P. Simonson, M. Sabuncu were with Weill Cornell Medicine, New York, NY, 10044, USA}}


\maketitle

\begin{abstract}
Continual learning is rapidly emerging as a key focus in computer vision, aiming to develop AI systems capable of continuous improvement, thereby enhancing their value and practicality in diverse real-world applications. In healthcare, continual learning holds great promise for continuously acquired digital pathology data, which is collected in hospitals on a daily basis. However, panoramic segmentation on digital whole slide images (WSIs) presents significant challenges, as it is often infeasible to obtain comprehensive annotations for all potential objects, spanning from coarse structures (e.g., regions and unit objects) to fine structures (e.g., cells). This results in temporally and partially annotated data, posing a major challenge in developing a holistic segmentation framework. Moreover, an ideal segmentation model should incorporate new phenotypes, unseen diseases, and diverse populations, making this task even more complex. In this paper, we introduce a novel and unified Incremental Relationship-guided Segmentation (IRS) learning scheme to address temporally acquired, partially annotated data while maintaining out-of-distribution (OOD) continual learning capacity in digital pathology. The key innovation of IRS lies in its ability to realize a new spatial-temporal OOD continual learning paradigm by mathematically modeling anatomical relationships between existing and newly introduced classes through a simple incremental universal proposition matrix. Experimental results demonstrate that the IRS method effectively handles the multi-scale nature of pathological segmentation, enabling precise kidney segmentation across various structures (regions, units, and cells) as well as OOD disease lesions at multiple magnifications. This capability significantly enhances domain generalization, making IRS a robust approach for real-world digital pathology applications.
\end{abstract}


\section{Introduction}
\label{sec:introduction}

\begin{figure}
\centering
\includegraphics[width=0.45\textwidth]{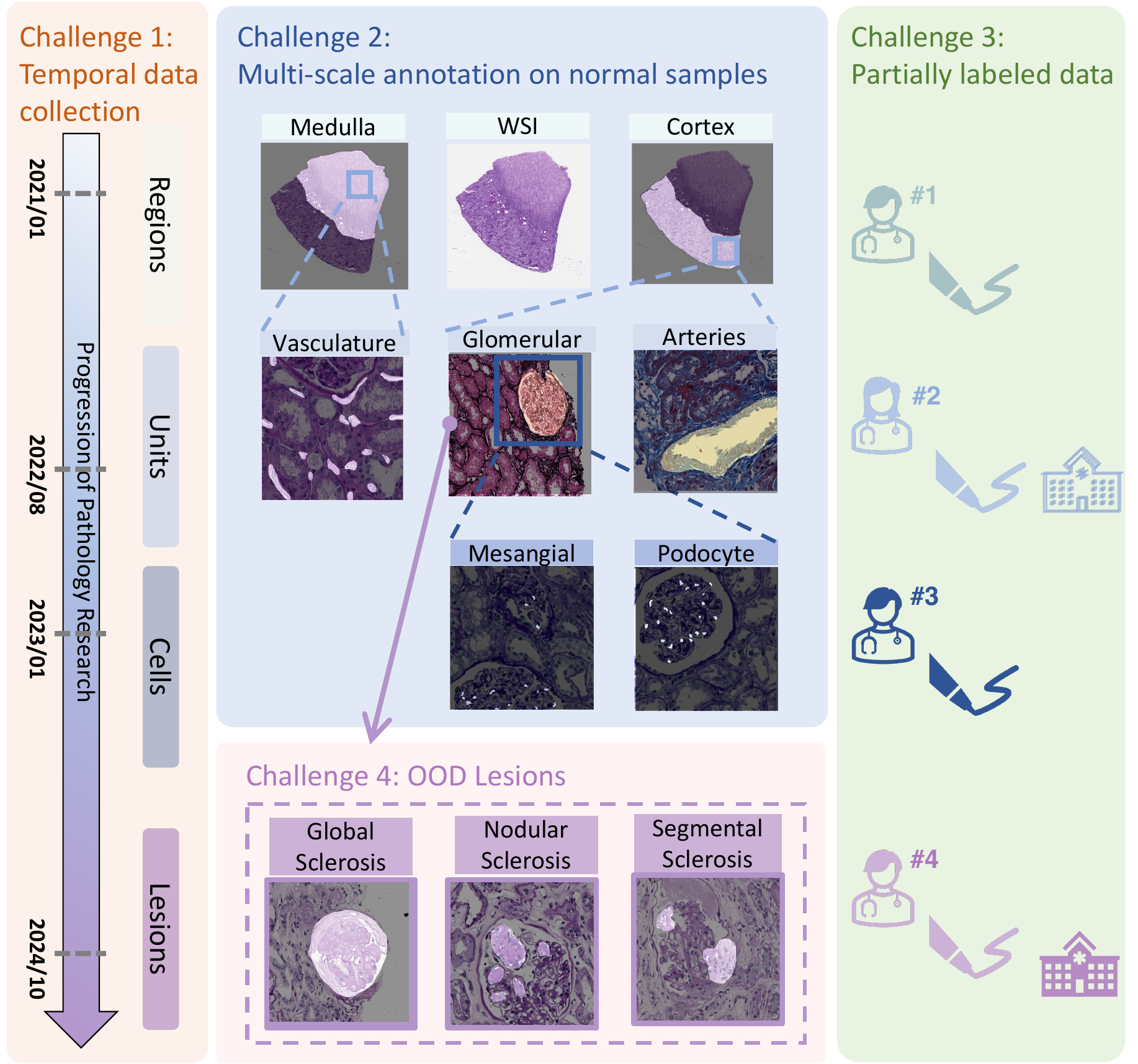}
\caption{Illustration of continual learning challenges in renal pathology segmentation, depicting: (1) temporal data collection; (2) multi-scale comprehensive objects ranging from regions to cells; (3) partial annotations from different annotators; and (4) progression from normal anatomical structures to out-of-distribution (OOD) pathological lesions.}
\label{fig.pipeline} 
\end{figure}

\begin{figure*}[t]
\centering 
\includegraphics[width=0.9\textwidth]{{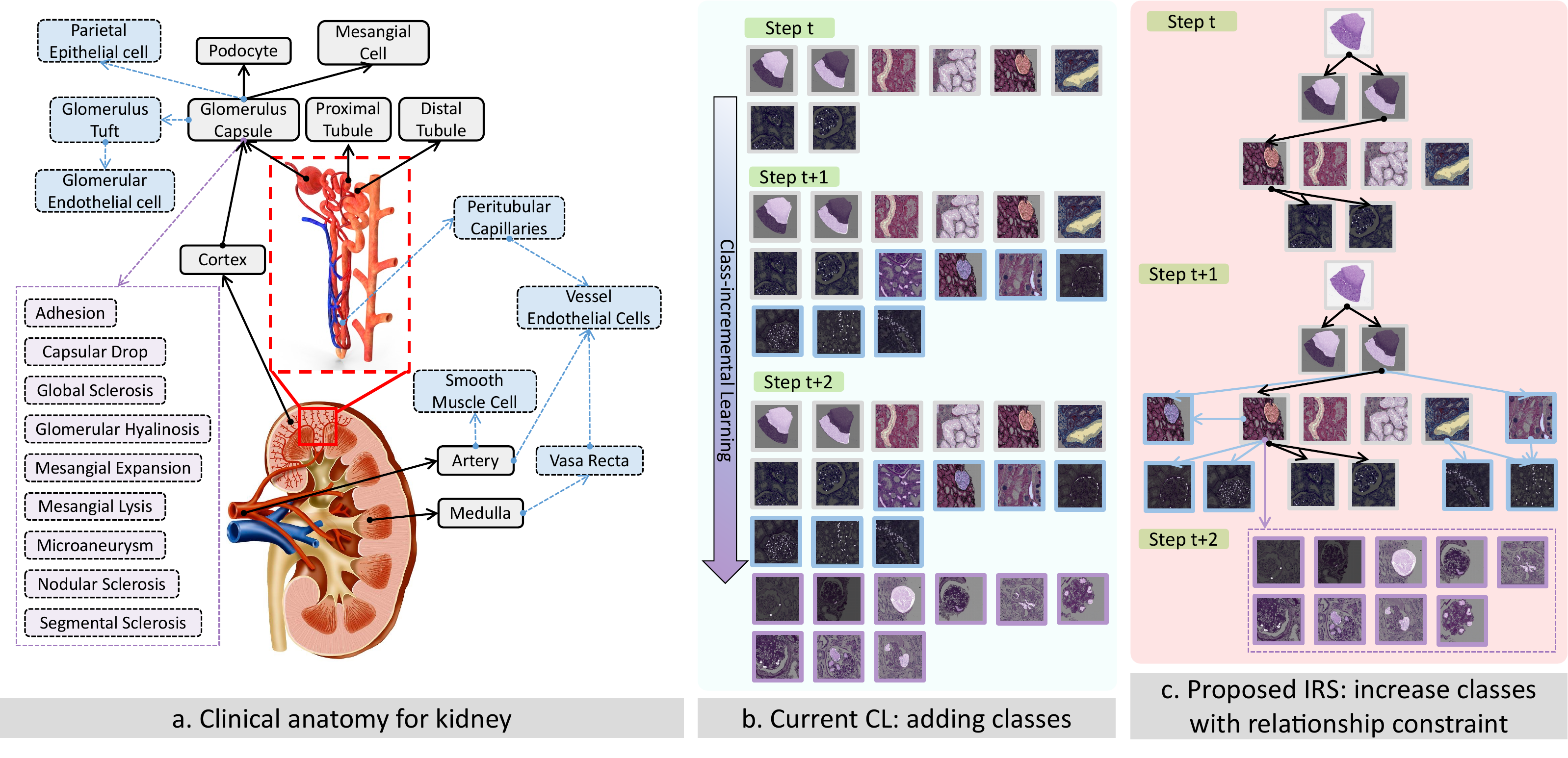}}
\caption{This figure illustrates the transformation of complex clinical anatomical relationships within the kidney into a continual learning paradigm. (a) The kidney's anatomy showcases the spatial relationships among large-scale objects, including regions, functional units, cells, and lesions. (b) Previous continual learning approaches merely add new classes without establishing connections between new and old classes. (c) The proposed IRS method integrates anatomical knowledge into continual learning to preserve knowledge from old classes, even when only new classes are used during model training.} 
\label{Fig:anatomy-awareness} 
\end{figure*}

\IEEEPARstart{C}{ontinual} learning is rapidly emerging as a key focus in computer vision, aiming to develop AI systems that can continuously adapt and improve~\cite{van2022three}. This ability to learn incrementally enhances their value and practicality across a wide range of real-world applications, particularly in clinical workflows within healthcare domains. By enabling models to incorporate new information without retraining from scratch or forgetting previously learned knowledge, continual learning addresses the limitations of traditional static models in scenarios such as task expansion~\cite{ruvolo2013ella,masse2018alleviating,ramesh2021model}, domain knowledge enlargement~\cite{ke2021classic,mirza2022efficient}, and class extension~\cite{von2019continual,shin2017continual,van2020brain}.

In the context of medical image analysis, the continual learning approach is especially valuable since medical data is inherently dynamic, with new data continuously collected~\cite{li2023systematic, manjunath2021systematic, vokinger2021continual}. For instance, digital whole slide images (WSIs) from new patients are generated daily, and segmentation models \emph{(1) must incorporate new phenotypes, previously unseen diseases, and diverse patient populations in line with evolving clinical guidelines}\cite{maher2019passive, kwok2022data, schobel2016towards}. Medical imaging modalities and protocols also evolve over time, introducing variations to which models must adapt. Specifically, achieving panoramic segmentation in renal pathology\cite{barisoni2020digital} requires annotations for numerous objects of interest spanning multiple scales and varying magnifications (as shown in Fig.~\ref{Fig:anatomy-awareness}), making it impractical to comprehensively annotate all structures simultaneously. This results in \emph{(2) temporally and partially annotated data}, creating a significant challenge for holistic segmentation. Moreover, there is a \emph{(3) progression from coarse structures (e.g., regions and unit objects) to fine structures (e.g., cells), and from normal anatomical structures to out-of-distribution (OOD) disease lesions}, necessitating continuous adaptation by segmentation models.

While deep learning techniques have advanced continual learning in pathological image analysis~\cite{perkonigg2021dynamic, zhang2023continual, chen2024low, gonzalez2023lifelong}, current continual learning pipelines generally treat the addition of new labels similarly to standard natural image analysis problems. Most of them focus merely on knowledge distillation from older models to retain previously learned classes (as shown in Fig.~\ref{Fig:anatomy-awareness}.b). In renal pathology, the extensive anatomical relationships among kidney structures (as shown in Fig.~\ref{Fig:anatomy-awareness}.a) provide valuable spatial correlations that can guide model learning, improving segmentation quality across diverse objects~\cite{li2022deep, ke2022unsupervised, deng2024prpseg}. These spatial-temporal relationships can facilitate incremental segmentation across multiple scales, transfer phenotype knowledge from existing classes to previously unseen classes, and bridge the gap between normal anatomical structures and OOD disease lesions.

In this paper, we introduce a novel and unified Incremental Relationship-guided Segmentation (IRS) learning scheme, designed specifically to manage temporally acquired, partially annotated data while preserving OOD continual learning capacity. Unlike traditional continual learning for natural images, which simply adds new classes, IRS leverages anatomical relationships between existing and newly introduced classes through a simple, yet powerful, incremental universal proposition matrix, mathematically modeling spatial-temporal relationships within digital pathology. We also propose a prompt-driven dynamic mixture-of-experts backbone, allowing the reuse of the same architectural framework across incremental steps, minimizing model modifications when new classes are introduced. The IRS model incorporates three backbones to achieve superior image feature embeddings for segmentation at multiple scales. Ultimately, IRS represents a single, scalable solution within a continual learning framework, achieving precise segmentation performance across various kidney structures (regions, units, cells) and OOD disease lesions, significantly enhancing domain generalization capabilities in real-world clinical applications. The official implementation is publicly available at \url{https://github.com/hrlblab/IRS}.

The contributions of this paper are threefold: \begin{itemize} \item We propose the IRS scheme that captures spatial-temporal relationships between existing and newly introduced classes by embedding clinical knowledge in a expandable incremental universal proposition matrix; \item A prompt-driven dynamic mixture-of-experts (MoE) model that ensures stability and adaptability as the model continues to learn; \item The proposed segmentation pipeline effectively addresses the multi-scale nature of pathological segmentation, enabling precise kidney segmentation across various structures (regions, units, cells) and OOD disease lesions at multiple magnifications, thereby promoting domain generalization. \end{itemize}

\section{Related Work}

\subsection{Partially Labeled Renal Pathology Segmentation}
\label{subsec:segmentation}
Recent advancements in deep learning have established Convolutional Neural Networks (CNNs) and Transformer-based networks as leading methods for image segmentation in renal pathology~\cite{feng2022artificial,hara2022evaluating,gadermayr1708cnn,gallego2018glomerulus,bueno2020glomerulosclerosis,lutnick2019integrated,gupta2018iterative,kannan2019segmentation}. Additionally, instance segmentation and Vision Transformers (ViTs) have found applications in this domain~\cite{johnson2019automatic,nguyen2021evaluating,gao2021instance,yan2023self,marechal2022automatic}.

While multi-head single network designs have been proposed for multi-class renal pathology segmentation~\cite{gonzalez2018multi, fang2020multi, bouteldja2021deep, chen2019med3d}, they often require dense multi-class annotations, which are labor-intensive. Given that pathology data is frequently partially labeled, recent developments in dynamic neural networks have enabled more comprehensive segmentation using single multi-label networks, even with incomplete data~\cite{zhang2021dodnet, deng2023omni}. However, comprehensive approaches capable of spanning from the tissue region level to the cellular level remain largely unexplored. Furthermore, some methods prioritize the segmentation of disease-positive regions over a holistic understanding of kidney morphology~\cite{jing2022segmentation, lin2022adversarial}.

This paper presents a dynamic mixture-of-experts model that unifies multiple backbones through self-attention and a dynamic head, capturing multi-scale features in renal pathology while ensuring stability and adaptability. Thus, it supports class-incremental learning for integrating partially labeled and OOD new classes.

\subsection{Continual Image Analysis}
\label{subsec:lifelonganalysis}

In medical data collection workflows, training data for AI models arrives sequentially, with distributions evolving over time~\cite{van2022three}. As new patient imaging data is generated daily, models must adapt to new phenotypes and updated clinical labels, aligning with evolving guidelines. Continual learning, therefore, requires computational efficiency and model sustainability. Continual learning in medical imaging has three main scenarios: (1) Task-incremental learning, where sequential data targets different tasks, as seen in multi-task learning (classification, segmentation, anomaly detection) across X-ray datasets~\cite{liao2022muscle}; (2) Domain-incremental learning, where models adapt to different domains to solve consistent problems, such as segmentation~\cite{butoi2023universeg} or classification~\cite{ye2024continual} across various data modalities.

The last category of continual learning, and the main problem we aim to tackle, is class-incremental learning, where the model is required to incorporate more data classes in different steps of training. Several continual learning approaches for medical image segmentation have achieved promising performance when new classes were introduced~\cite{ji2023continual, zhang2023continual, gonzalez2023lifelong}. However, these methods merely include more classes and provide semi-supervised learning strategies between different steps to retain knowledge from previous classes, while the model architectures are normally modified to accommodate new classes. 

In our IRS, we integrate spatial-temporal relationships between existing and new classes into continual learning using an expandable incremental universal proposition matrix. An anatomical loss supervises spatial-temporal relationships, distilling prior class knowledge.

\section{Method}
\label{sec:method}


We address the task of continual learning segmentation, formulated as \textbf{class-incremental semantic segmentation}, in the context of kidney pathology (Fig.~\ref{Fig:anatomy-awareness}). Our aim is to develop a model that can continuously learn to segment new anatomical structures and lesions over multiple steps during temporal data collection in clinical workflows, while retaining knowledge of previously learned classes.




\begin{figure}[t]
\centering 
\includegraphics[width=0.9\linewidth]{{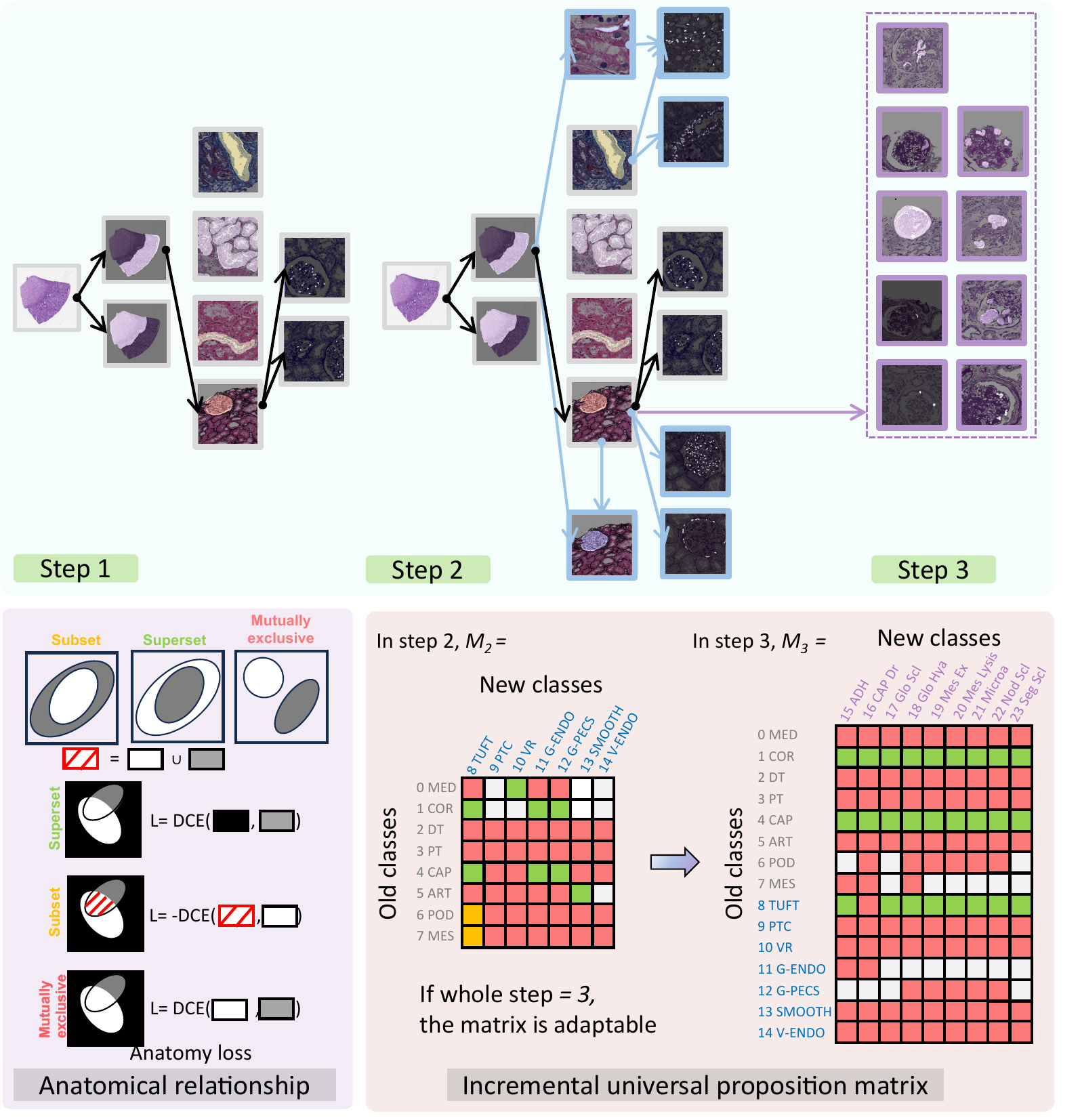}}
\caption{This figure illustrates the design of the incremental universal proposition matrix as new classes are added in different settings. The proposed matrix is easily adaptable to large-scale objects within the continual learning paradigm.}
\label{Fig.flexibility} 
\end{figure}



\subsection{Incremental Universal Proposition Matrix}
\label{subsec:matrix}

Renal pathology objects follow anatomical flows through heterogeneous structures. In continual learning, adding classes with relational information enhances analysis efficiency. Inspired by~\cite{deng2024prpseg}, we model clinical anatomy concepts as a mathematical framework to represent the relationships between old and new classes.

Building upon~\cite{deng2024prpseg}, we introduce a incremental universal proposition matrix, \( M_t \in \mathbb{R}^{m \times n} \), specifically designed to bridge the relationships between old and new classes. Here, \( m \) represents the number of old classes, and \( n \) represents the number of new classes. The matrix values are defined in~\cite{deng2024prpseg} for \( i = 1,\dots,m \) and \( j = 1,\dots,n \). This matrix is flexible and adaptable to different class settings in continual learning as sequential data is introduced, as shown in Fig.~\ref{Fig.flexibility}.

At step \( t \), the previous classes \( C_{1:t-1} \) are combined with the new classes \( C_t \) to form \( C_{1:t} \). At step \( t+1 \), the classes \( C_t \) are added to the old classes to form \( C_{1:t+1} \), and the matrix is rebuilt to incorporate the new classes.

\subsection{Prompt-Driven Dynamic Mixture-of-Experts Network}
\label{subsec:token}

\begin{figure}
\centering 
\includegraphics[width=0.9\linewidth]{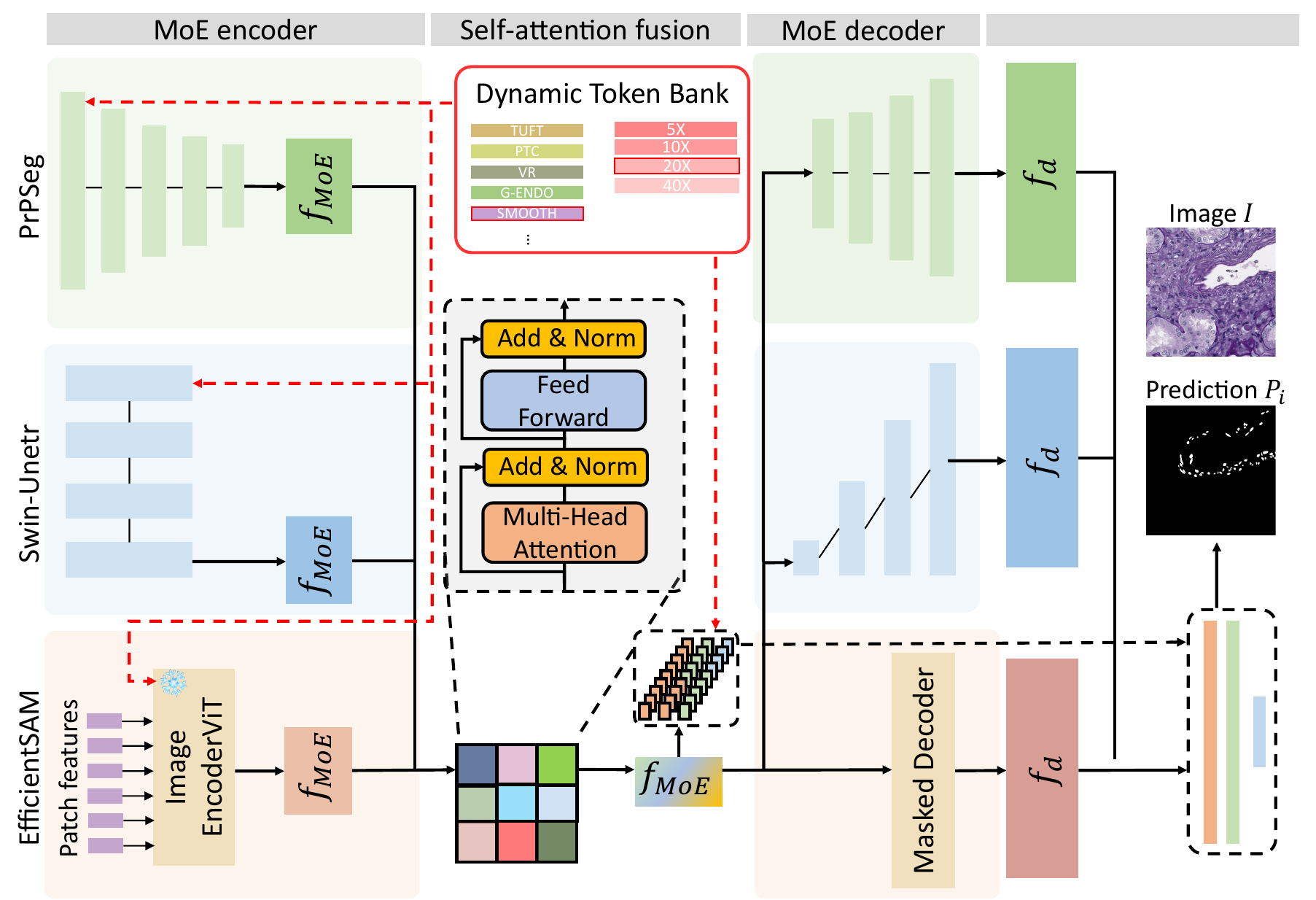}
\caption{This figure illustrates the architecture of our proposed prompt-driven dynamic MoE network. The model integrates three different architectures using a self-attention mechanism and a dynamic MoE head to enhance segmentation capabilities for large-scale pathology segmentation in continual learning. Additionally, the model maintains a consistent architecture while dynamically accommodating an increasing number of segmentation classes.
} 
\label{Fig.token} 
\end{figure}
Incremental pathological image segmentation networks face two challenges: varying segmentation backbones limit comprehensive multi-scale segmentation, and multi-network approaches require extensive modifications for new classes. To address this, we propose a prompt-driven MoE network that enhances large-scale segmentation in continual learning while maintaining a stable, adaptable architecture (as shown in Fig.~\ref{Fig.token}).

\noindent\textbf{Mixture-of-Experts Backbone:} The network incorporates three different backbones that have shown superior segmentation performance on various medical objects to satisfy the demands of increasing class numbers. The backbone of the proposed network integrates architectures from PrPSeg~\cite{deng2024prpseg}, Swin-UNETR~\cite{hatamizadeh2021swin}, and EfficientSAM~\cite{xiong2023efficientsam}, which have demonstrated superior segmentation performance for multi-scale objects across regions, functional units, and cells. Each image $I$ is input to multiple encoders $F_e^n$ of different experts to obtain latent features $f_{\text{latent}}^n$. A self-attention mechanism is implemented to fuse these different embeddings into the MoE latent feature $f_{\text{MoE}}$, defined by the following algorithm:

\begin{algorithm}
\caption{Self-Attention Module}
\KwIn{Input image $I$, initial feature map $X$}
\KwOut{MoE latent feature $f_{\text{MoE}}$}

\For{$n \gets 1$ \textbf{to} $3$}{
    $f_{\text{latent}}^n \leftarrow F_{e}^n(I)$\;
}
$A \leftarrow \text{MultiHeadAttention}(f_{\text{latent}}^1, f_{\text{latent}}^2, f_{\text{latent}}^3)$\;
$X' \leftarrow X + \text{Dropout}(A)$\;
$X'' \leftarrow \text{BatchNorm1d}(X'^\top)^\top$\;
$F \leftarrow \text{FeedForward}(X'')$\;
$X''' \leftarrow X'' + \text{Dropout}(F)$\;
$f_{\text{MoE}} \leftarrow \text{BatchNorm1d}(X'''^\top)^\top$\;
\label{alg:selfattention}
\end{algorithm}

\noindent\textbf{Learnable Dynamic Token Bank:} A predefined learnable token bank is initialized to store the class-specific and scale-specific knowledge across the whole dataset. Dimensionally stable class-aware tokens ($T_c \in \mathbb{R}^{k \times d}$) and scale-aware tokens ($T_s \in \mathbb{R}^{4 \times d}$) are employed from the token bank to capture contextual information in the model. Each class has a one-dimensional token, $t_c \in \mathbb{R}^{1 \times d}$, to store class-specific knowledge at the feature level across the entire dataset, while each magnification scale has a one-dimensional token, $t_s \in \mathbb{R}^{1 \times d}$, to provide scale-specific knowledge across four scales ($5\times$, $10\times$, $20\times$, and $40\times$).

There are two modules that use these conditional tokens to achieve semantic segmentation in the network: (1) Inspired by the Vision Transformer (ViT)~\cite{dosovitskiy2020image}, for an image $I$ of class $i$ with magnification $m$, the corresponding class token $T_c[i]$ and scale token $T_s[m]$ are stacked with the patch-wise image tokens before being fed into each block of MoE encoders ($F_e^n$), as shown in~\eqref{eq:encodertoken}, where $n = 1,2,3$ in this study; (2) to achieve binary segmentation for each class (due to partially labeled datasets in the medical field), $T_c[i]$ and $T_s[m]$ are concatenated with the MoE latent feature $f_{\text{MoE}}$ to produce the parameters ($\omega$) in the dynamic head, as shown in~\eqref{eq:fusion}.

\begin{equation}
\begin{aligned}
    f_e^n = F_e^n(T_c[i] \, || \, T_s[m] \, || \, e_{b-1})
\end{aligned}
\label{eq:encodertoken}
\end{equation}

\begin{equation}
    \omega = \varphi(f_{\text{MoE}} \, || \, T_c[i] \, || \, T_s[m]; \Theta_\varphi)
\label{eq:fusion}
\end{equation}

\noindent where $||$ represents the stacking operation, and $\Theta_\varphi$ denotes the parameters in the dynamic head. 

\noindent\textbf{MoE Dynamic Head:} Following the approach of~\cite{deng2023omni}, a binary segmentation network is used to achieve multi-label segmentation via a dynamic filter incorporating heterogeneous information from MoE decoder features ($F_d^n$). The dynamic head consists of three layers. The first two layers contain eight channels each, while the final layer comprises two channels. We directly map parameters from the fusion-based feature controller to the kernels in the 162-parameter dynamic head to achieve precise segmentation from multi-modal features. The filtering process is expressed by the following equations:

\begin{equation}
\begin{aligned}
   & f_d = \sum_{n=1}^{3} F_d^{n}(f_{\text{MoE}})\\
   & P = (((f_d \cdot \omega_1) \cdot \omega_2) \cdot \omega_3)
\end{aligned}
\label{eq:dynamic-head}
\end{equation}

\noindent where $\cdot$ denotes convolution, $P \in \mathbb{R}^{2 \times W \times H}$ is the final prediction, and $W$ and $H$ correspond to the width and height of the dataset, respectively.

\subsection{Relationship-Guided Knowledge Distillation}
\label{subsec:knowledgedistillation}

Since continual learning does not include labels for old classes in each step, this leads to catastrophic forgetting of old classes. Classical continual learning strategies merely add new classes without any cooperation between classes, focusing on knowledge distillation from the previous model to the current model. However, the extensive anatomical relationships existing among panoramic objects are critical for guiding the model in remembering knowledge from old classes and understanding knowledge for new classes in renal pathology.

With the introduction of the incremental universal proposition matrix, we are able to supervise the knowledge of old classes by incorporating spatial correlations between old class predictions and new class labels into the training process (as shown in Fig.~\ref{Fig.flexibility}). At the current stage $t$, for a given image $I$ from the new classes ($C_t$) with a labeled class $j$, represented as $Y_j$, we generate predictions $P_i$ for another class $i$ from the old classes ($C_{1:t-1}$) on the same image. We then use the anatomical relationships to constrain the correlation between the supervised label $Y_j$ and the semi-supervised prediction $P_i$:

\begin{enumerate}
    \item If $j$ is a superset of $i$, $P_i$ should not exceed region $Y_j$.
    \item If $j$ is a subset of $i$, $P_i$ should cover $Y_j$.
    \item If $j$ is mutually exclusive with $i$, the overlap between $Y_j$ and $P_i$ should be minimized.
\end{enumerate}

The total anatomy loss is defined by: 
\begin{equation}
L_{\text{anatomy}}(j,i) = 
\begin{cases}
\text{DCE}(1 - Y_j, P_i), & \text{if } j \triangleright i \\
- \text{DCE}(Y_j, Y_j \cup P_i), & \text{if } j \triangleleft i \\
\text{DCE}(Y_j, P_i), & \text{if } j \parallel i \\
0, & \text{otherwise}
\end{cases}
\label{eq:taxonomyloss}
\end{equation}

\noindent where $
m = M_t(i,j), \quad i \in C_{1:t-1}, \quad j \in C_t$. $M_t$ is the incremental matrix at step $t$, and $\text{DCE}$ denotes the Dice Loss. With this relationship-guided semi-supervised constraint, the model is able to remember previous classes by using the new labels.

\begin{figure}
\centering 
\includegraphics[width=0.9\linewidth]{{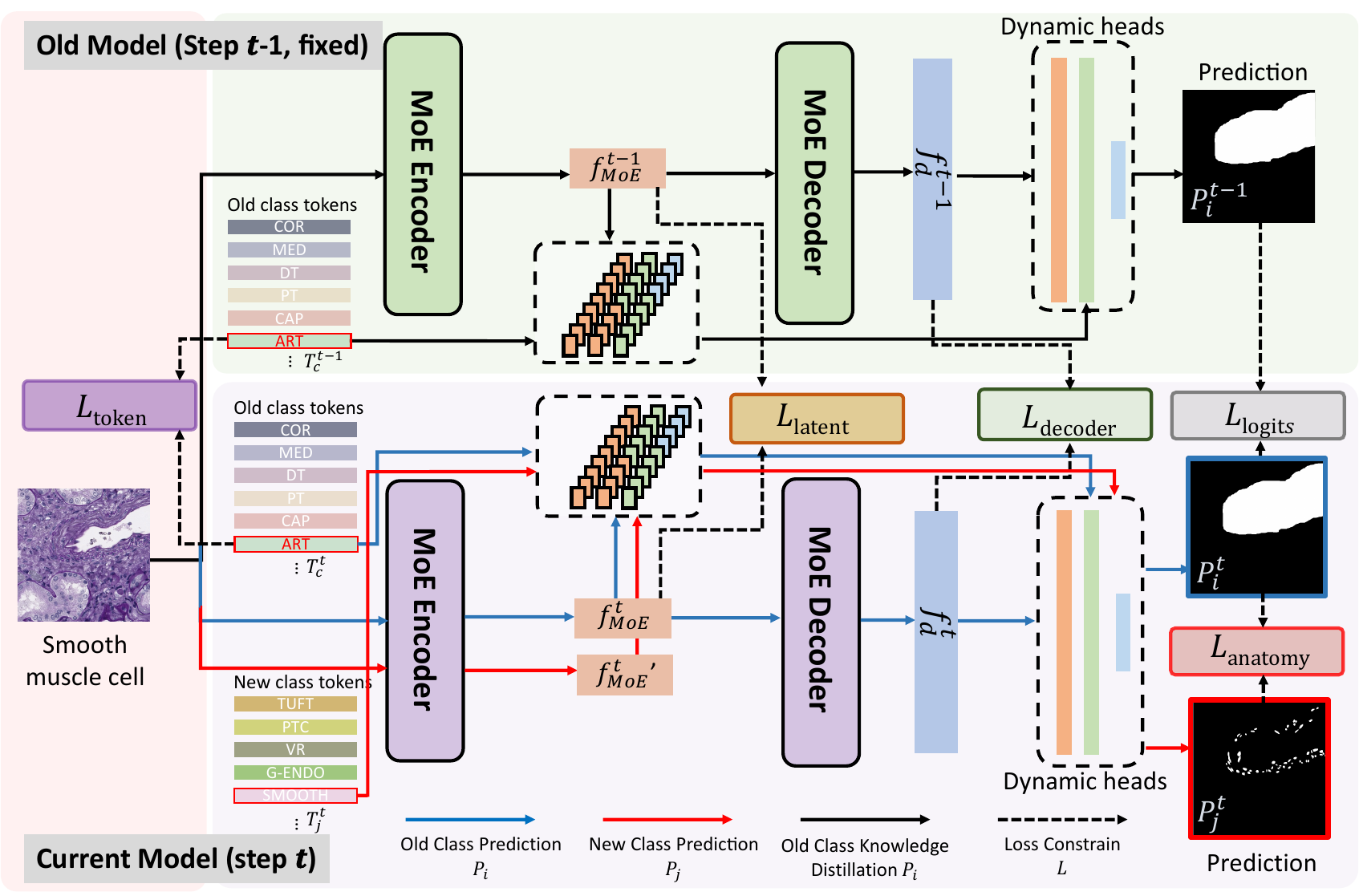}}
\caption{This figure showcases the key innovation of knowledge distillation in our proposed method. In subsequent steps, the model learns from new classes supervised by new labels, while old class tokens are utilized on new images to predict old classes. The new model distills knowledge from the previous model by maintaining similarity in old class tokens, latent features, decoder features, and prediction logits.}
\label{fig.anatomy} 
\end{figure}

At the same time, in order to distill the class-aware knowledge from the previous classes ($C_{1:t-1}$) without using their labels ($Y_i$), we incorporate the model from the previous step $t-1$ and calculate the consistency among different levels of features and outputs ($f_{\text{MoE}}$, $f_d$, $P_i$) from the model (as shown in Fig.~\ref{fig.anatomy}). Since we designed individual class tokens to represent the features for each class, the $T_c[i]$ of previous classes ($C_{1:t-1}$) should remain consistent between different steps. Therefore, we combine similarity losses implemented with KL divergence loss and MSE loss, and define the distillation loss as:

\begin{equation}
\begin{aligned}
& L_{\text{consistency}}(a,b) = \text{KL\_Loss}(a,b) + \text{MSE\_Loss}(a,b) \\
& \text{for } i \in C_{1:t-1}: \\
& \quad L_{\text{token}}(i,t-1,t) = L_{\text{consistency}}(T_c[i]^{t-1}, T_c[i]^{t}) \\
& \quad L_{\text{latent}}(i,t-1,t) = L_{\text{consistency}}(f_{\text{MoE}}^{t-1}, f_{\text{MoE}}^{t}) \\
& \quad L_{\text{decoder}}(i,t-1,t) = L_{\text{consistency}}(f_d^{t-1}, f_d^{t}) \\
& \quad L_{\text{logits}}(i,t-1,t) = L_{\text{consistency}}(P_i^{t-1}, P_i^{t}) \\
& L_{\text{semi}} = L_{\text{token}} + L_{\text{latent}} + L_{\text{decoder}} + L_{\text{logits}}
\end{aligned}
\label{eq:semi}
\end{equation}

The total loss function is an aggregate of supervised loss using images and labels from the current step and anatomy-aware semi-supervised losses involving classes from previous steps, enabling the learning of new classes while retaining knowledge of old classes in continual learning.

\section{Data and Experiment}
\label{sec:details}

\begin{table}[htbp]
\centering
\caption{Data collection.}
\begin{adjustbox}{width=0.45\textwidth}
\begin{tabular}{ll|cccc}
\toprule
Class name & Abbreviation  & Patch \#  & Size & Scale & Stain\\
\midrule
Medulla & Medulla & 1619 &  1024$^{2}$ & 5 $\times$ & P \\
Cortex & Cortex & 3055 & 1024$^{2}$ & 5 $\times$ & P\\
\midrule
Distal tubular & DT & 4615 & 256$^{2}$ & 10 $\times$ & H,P,S,T \\
Proximal tubular &  PT & 4588 &  256$^{2}$ & 10 $\times$ & H,P,S,T \\
Glomerular capsule & Cap. & 4559  &  256$^{2}$ & 5 $\times$ & H,P,S,T  \\
Glomerular tufts & Tuft &  4536 &  256$^{2}$ & 5 $\times$ & H,P,S,T \\
Arteries & Art. & 4875 &  256$^{2}$ & 10 $\times$ & H,P,T \\
Peritubular capillaries & PTC & 4827  &  256$^{2}$ & 40 $\times$ & P\\
Vasa Recta & VR & 1362  &  512$^{2}$ & 20 $\times$ & P\\
\midrule
Podocytes & Pod. & 1147  & 512$^{2}$ & 20 $\times$ & P \\
Mesangial cells & Mes. & 789  & 512$^{2}$ & 20 $\times$ & P \\
Glomerular endothelial cells & G-Endo. & 715 & 512$^{2}$ & 20 $\times$ & P  \\
Parietal epithelial cells & G-Pecs & 2014  & 512$^{2}$ & 20 $\times$ & P \\
Smooth muscle cells & Smooth. & 1326  & 512$^{2}$ & 20 $\times$ & P \\
Vessel endothelial cells & V-Endo. & 1304  & 512$^{2}$ & 20 $\times$ & P\\
\midrule
Tuft adhesion & ADH & 72  & 512$^{2}$ & 20 $\times$ & P\\
Capsular droplet & Cap. Dr. & 16 & 512$^{2}$ & 20 $\times$ & P\\
Global sclerosis & Glo. Scl. & 2960 & 512$^{2}$ & 20 $\times$ & P\\
Hyalinosis & Glo. Hya. & 288 & 512$^{2}$ & 20 $\times$ & P\\
Mesangial expansion & Mes. Ex. & 1888 & 512$^{2}$ & 20 $\times$ & P\\
Mesangial lysis & Mes. Lysis. & 184 & 512$^{2}$ & 20 $\times$ & P\\
Microaneurysm & Microa. & 448 & 512$^{2}$ & 20 $\times$ & P \\
Nodular sclerosis & Nod. Scl. & 1872 & 512$^{2}$ & 20 $\times$ & P \\
Segmental glomerular sclerosis & Seg. Scl.  & 560 & 512$^{2}$ & 20 $\times$ & P \\
\bottomrule
\end{tabular}
\end{adjustbox}
\text{*H is H$\&$E; P is PAS; S is SIL; T is TRI.}
\label{tab:dataset} 
\end{table}

\begin{table*}[htbp]
\centering
\caption{The performance of continual learning in a 2-step setting, following the order $D_1$ and then $D_2$, is reported using Dice similarity coefficient scores (\%).
}
\begin{adjustbox}{width=0.7\textwidth}
\begin{tabular}{l|ccccccccc|c}
\toprule
\multirow{2}{0.8in}{Method} & \multicolumn{3}{c}{Old Classes} & \multicolumn{3}{c}{New Classes} & \multicolumn{3}{c}{Average} & Statisic\\
\cmidrule(lr){2-4}
\cmidrule(lr){5-7}
\cmidrule(lr){8-10}
 & Regions & Units & Cells & Units  & Cells  & Lesions & Old  & New  & All & \\
\midrule
PLOP~\cite{douillard2021plop} & 64.40 & 44.39 & 49.55 & 49.09 & 51.17 & 52.61 & 50.68 & 51.59 & 51.29 & $p <$ 0.001 \\
MiB~\cite{cermelli2020modeling} & 60.13 & 55.14 & \textbf{55.78} & 48.26 & 49.80 & 52.06 & 56.55 & 50.78 & 52.70 & $p <$ 0.001\\
Incrementer~\cite{shang2023incrementer} & 23.01 & 51.42 & 50.59 & 71.50 & 59.98 & 59.58 & 44.11 & 61.92 & 55.03 & $p <$ 0.001\\
CoMFormer~\cite{cermelli2023comformer} & 23.01 & 51.42 & 50.59 & 71.50 & 59.98 & 59.58 & 44.11 & 61.92 & 55.98 & $p <$ 0.001 \\
ILTS~\cite{michieli2019incremental} & 65.42 & 54.07 & 49.88 & 64.62 & 56.72 & 56.93 & 55.86 & 58.32 & 57.50 & $p <$ 0.001\\
REMINDER~\cite{phan2022class} & 58.21 & 57.53 & 49.85 & 61.22 & 49.80 & 49.65 & 55.78 & 51.86 & 53.16 & $p <$ 0.001 \\
\midrule
CLIP-CL~\cite{zhang2023continual} & 22.90 & 44.55 &  50.18 &  71.74 & 59.33 & 59.06 & 40.54 & 61.51 & 54.52 & $p <$ 0.001\\
CL-LoRA~\cite{chen2024low} & \textbf{67.02} & 54.25 & 52.73 &  66.05 & 58.10 & 54.82 & 57.06 & 57.74 & 57.52 & $p <$ 0.001\\
\midrule
IRS (Ours) & 66.53 &  \textbf{57.97} & 50.02 & \textbf{73.62} &  \textbf{64.03} &  59.80 &  \textbf{58.13} & \textbf{63.44} & \textbf{61.67} & \textbf{Ref.}\\
\midrule
\rowcolor{gray!20} 
Joint & 71.68 & 73.07 & 63.61 & 73.37 & 60.47 & 58.96 & 70.36 & 62.04 & 64.81 & $p <$ 0.001\\
\bottomrule
\end{tabular}
\end{adjustbox}
\label{tab:2step_Qualitative}
\end{table*}

\begin{figure*}
\centering 
\includegraphics[width=0.8\linewidth]{{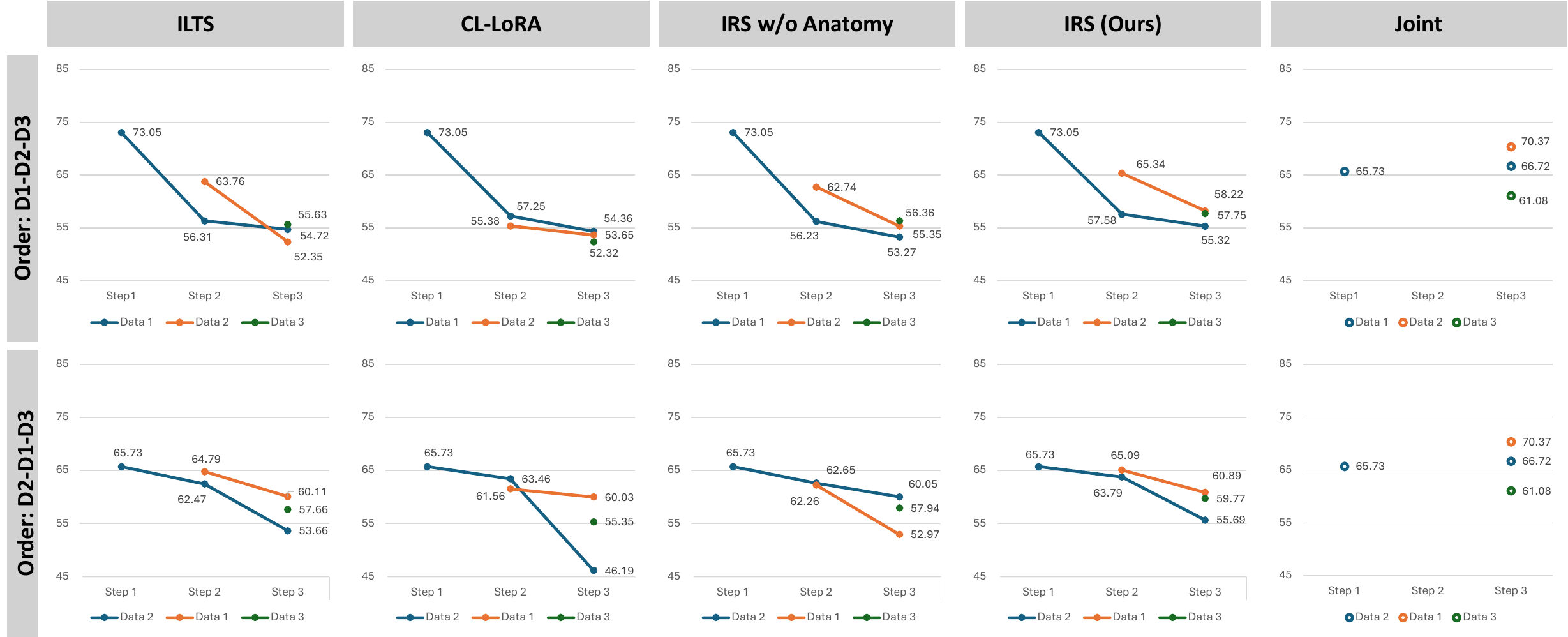}} \caption{This figure shows the results of 3-step continual learning with 2 orders. Dice similarity coefficient scores (\%) are reported. The proposed method achieved superior performance in panoramic renal pathology segmentation across different continual learning settings. } 
\label{Fig.Qualitative_3step} 
\end{figure*}

\begin{table*}[htbp]
\centering
\caption{Performance of continual learning in a single-class 4-step setting, following the order of Cortex-Capsule-Podocyte-Nodular sclerosis. Dice similarity coefficient scores (\%) are reported.}

\begin{adjustbox}{width=0.7\textwidth}
\begin{tabular}{l|c|cc|ccc|cccc|ccc}
\toprule
\multirow{2}{0.8in}{Method} & \multicolumn{1}{c}{S1} & \multicolumn{2}{c}{S2} & \multicolumn{3}{c}{S3} & \multicolumn{4}{c}{S4} & S2 Ave. & S3 Ave. & S4 Ave.\\
 & Cor. &  Cor.  & Cap. & Cor. & Cap. & Pod.  & Cor. & Cap. & Pod. & Nod. Scl. & & &  \\
\midrule
ILTS~\cite{michieli2019incremental} & 72.05&70.36&84.99&\textbf{70.86}&48.64&\textbf{63.85}&50.53&50.53&50.53&50.53&77.67&61.12&50.53 \\
CL-LoRA~\cite{chen2024low} & 72.05&71.17&74.98&69.31&45.48&50.48&69.65&45.43&50.48&49.31&73.08&55.09&53.72 \\
\midrule
IRS w/o A. &72.05&71.85&88.56&70.62&52.44&71.12&43.14&\textbf{56.42}&50.23&\textbf{75.40}&80.21&64.73&56.30\\
IRS (Ours) & 72.05&\textbf{72.16}&\textbf{89.50}&70.07&\textbf{72.79}&62.85&\textbf{70.62}&52.44&\textbf{66.04}&53.49&\textbf{80.83}&\textbf{68.57}&\textbf{60.65} \\
\midrule
\rowcolor{gray!20} 
Joint & 72.05 &-&-&-&-&-&72.87&88.46&63.40&64.49&-&-&72.31 \\
\bottomrule
\end{tabular}
\end{adjustbox} 
\label{tab:4-step}
\end{table*}

\subsection{Data}
\label{subsec:data}

Our model leverages a 24-class, partially labeled dataset spanning various biological scales. The dataset's structure and collection details are provided in Table~\ref{tab:dataset}. We sourced the human kidney dataset from four distinct resources, each originating from a different institution or clinical project, encompassing regions, functional units, cells, and OOD lesions. The dataset was partitioned into training, validation, and testing sets at a 6:1:3 ratio across all classes, with splits conducted at the patient level to prevent data leakage.

\subsection{Training Process}
\label{subsec:experiment}

To evaluate IRS, we employ 3 continual learning settings:

\noindent\textbf{2-step continual learning}: Two groups of data ($D_1$ and $D_2$) were organized according to the timestamps of data collection. $D_1$ includes labels for regions (medulla, cortex), functional units (distal/proximal tubules, glomerular capsule, artery), and cells (podocytes, mesangial cells). Additional units, cell types, and OOD lesions are introduced in $D_2$.

\noindent\textbf{3-step continual learning}: Three groups of data ($D_1$, $D_2$, and $D_3$) were organized according to the timestamps of data collection. $D_1$ includes labels for regions (medulla, cortex), functional units (distal/proximal tubules, glomerular capsule, artery), and cells (podocytes, mesangial cells). Additional units and cell types are introduced in $D_2$, while $D_3$ includes all OOD lesion samples.

\noindent\textbf{4-step continual learning}: Based on the principles of renal pathology, which progress from coarse (e.g., cortex, capsule) to fine (e.g., podocyte) structures, and from normal to lesional objects (e.g., nodular sclerosis), four classes are introduced sequentially in each epoch of continual learning.

\subsection{Experiment Details}
\label{subsec:experiment}

Our model’s training process consisted of two phases. In the first phase, we used supervised learning for 100 epochs on dataset $D_1$ to minimize both the binary Dice loss and the cross-entropy loss, allowing the model to grasp multi-scale object concepts. In subsequent steps, training continued for just one epoch per new class. 

Images were preprocessed to a uniform size of $512 \times 512$ pixels using random cropping/padding during training and center-cropping or tiling during testing. Backbone performance was initially evaluated on dataset $D_1$, using Adam optimizer (learning rate 0.001, decay 0.99), with augmentations such as affine transformations, contrast adjustments, and Gaussian noise. Model selection was based on mean Dice score on dataset $D_1$. Continual learning strategies used consistent hyperparameters from the initial training. All experiments ran on an NVIDIA RTX A6000 GPU for uniformity.

\begin{table*}
\caption{Ablation study of knowledge distillation on 24-class segmentation. Dice similarity coefficient scores (\%) are reported.}
\begin{center}
\begin{adjustbox}{width=0.7\textwidth}
\begin{tabular}{lcccccc|ccc|c}
\hline
Backbone & Freeze & $L_\text{token}$ & $L_\text{latent}$ & $L_\text{decoder}$ & $L_\text{logits}$ & $L_\text{anatomy}$ & Old Class & New Class & Average & Statistic.\\
\hline
MoE-3 & -- &  & \checkmark & \checkmark & \checkmark &  & 56.25 & 61.54 & 59.78 & $p <$ 0.001 \\
MoE-3 & -- & \checkmark & \checkmark & \checkmark & \checkmark &  & 56.56 & 62.17 & 60.30 & $p <$ 0.001 \\
\hline
MoE-3 & Fully & \checkmark & \checkmark & \checkmark & \checkmark & & 57.72 & 59.18 & 58.70 & $p <$ 0.001 \\
MoE-3 & Encoder & \checkmark & \checkmark & \checkmark & \checkmark & & 57.11 & 61.17 & 59.82 & $p <$ 0.001 \\
\hline
MoE-3 & -- & \checkmark & \checkmark &  & \checkmark & \checkmark & 58.02 & 62.67 & 61.12 & $p <$ 0.001\\
MoE-3 & -- & \checkmark & \checkmark &  & Psuedo-labels & \checkmark & 46.18 & 63.36 & 57.64 & $p <$ 0.001\\
MoE-3 (Ours) & -- & \checkmark & \checkmark & \checkmark & \checkmark & \checkmark & \textbf{58.13} & \textbf{63.44} & \textbf{61.67} & \textbf{Ref.} \\
\hline
\end{tabular}
\end{adjustbox}
\end{center}
\begin{tablenotes}
      \small
      \item *``MoE-3'' refers to a Mixture-of-Experts model with 3 backbones. ``Fully'' freezes all backbone weights, keeping only dynamic tokens and heads trainable. ``Psuedo-labels'' uses binary segmentation pseudo labels in the loss function instead of prediction logits.
    \end{tablenotes}
\label{table:semidesign} 
\end{table*}

\section{Results}
\label{sec:results}

We conducted a comparative analysis of our proposed IRS approach against various continual learning baseline models, which have achieved superior performance on natural image datasets~\cite{douillard2021plop,cermelli2020modeling,shang2023incrementer,cermelli2023comformer,michieli2019incremental,phan2022class} and medical image datasets~\cite{zhang2023continual,chen2024low}. Additionally, we include results from non-continual learning setting (where all class labels are provided during training), denoted as ``Joint” in the table for comparison.

\begin{table}
\caption{Ablation study of backbone design on dataset $D_1$ segmentation at step 1. Dice similarity coefficient scores (\%) are reported.}
\begin{center}
\begin{adjustbox}{width=0.45\textwidth}
\begin{tabular}{lccccccc}
\hline
Backbone & MoE & Token & Regions & Units & Cells & Average & Statistic.\\
\hline
Lifelong nnU-Net~\cite{gonzalez2023lifelong}& & & 44.95 & 75.66 & 58.12 & 63.59 & $p <$ 0.001\\
PrPSeg~\cite{deng2024prpseg}& & & 67.88 & 74.24 & 65.05 & 70.35 & $p <$ 0.001\\
Swin-Unetr~\cite{hatamizadeh2021swin} & & & 40.98 & 73.84 & 58.76 & 61.85 & $p <$ 0.001 \\
EfficientSAM~\cite{xiong2023efficientsam} & & & 71.40 & 75.59 & 60.24 & 70.70 & $p <$ 0.001 \\
\hline
MoE-3 & S.A. &  & 72.77 & 73.90 & 64.23 & 71.20 & $p <$ 0.001 \\
MoE-3 & S.A. & S.A. & 58.65 & 44.55 & 49.90 & 49.41 & $p <$ 0.001 \\
\hline
MoE-3 & Add & \checkmark & 71.65 & 74.97 & 65.46 & 71.76 & $p <$ 0.001 \\
MoE-5 & S.A. & \checkmark & 71.07 & \textbf{76.85} & 61.37 & 71.53 & $p <$ 0.001 \\
MoE-3 (Ours) & S.A. & \checkmark & \textbf{72.85} & 76.52 & \textbf{66.32} & \textbf{73.05} & \textbf{Ref.} \\
\hline
\end{tabular}
\end{adjustbox}
\end{center}

\begin{tablenotes}
      \small
      \item *MoE-3 is Mixture-of-Expert with 3 backones. MoE-5 is Mixture-of-Expert with 5 backones (adding UNI~\cite{chen2024uni}, and Prov-GigaPath~\cite{xu2024whole} models). S.A. is Self-attention.
    \end{tablenotes}

\label{table:backbonedesign} 
\end{table}

\subsection{2-Step Continual Segmentation Performance}
\label{subsec:2step_results}

Table~\ref{tab:2step_Qualitative} showcases the results from a 24-class continual segmentation evaluation in a 2-step setting, following the order $D_1$ and then $D_2$. The Dice similarity coefficient (Dice: \%, higher is better) was employed as the primary metric for quantitative performance assessment. The result demonstrates that our proposed method, surpasses baseline models in most evaluated metrics, showing that the proposed IRS achieves better segmentation performance across all old and new classes. Class-wise comprehensively quantitative results and qualitative results for all 24 classes can be found in Fig.~\ref{Fig.24classResults}.



\begin{table}
\centering
\caption{Performance of continual learning in a 2-step setting, following the order of $D_2$ and then $D_1$. Dice similarity coefficient scores (\%) are reported.
}
\begin{adjustbox}{width=0.45\textwidth}
\begin{tabular}{l|ccc|ccc|ccc|c}
\toprule
\multirow{2}{0.8in}{Method} & \multicolumn{3}{c}{Old Classes} & \multicolumn{3}{c}{New Classes} & \multicolumn{3}{c}{Average}\\
& Units  & Cells  & Lesions & Regions & Units & Cells & Old  & New  & All & Statistic. \\
\midrule
ILTS~\cite{michieli2019incremental} & 69.02&55.68&52.90&63.47&68.50&52.82&56.61&63.32&58.85 & $p <$ 0.001 \\
CL-LoRA~\cite{chen2024low} & \textbf{70.12}&\textbf{57.52}&56.27&64.08&65.22&50.13&59.18&61.16&59.84 & $p <$ 0.001\\
\midrule
IRS w/o anatomy & 69.61&54.42&60.30&\textbf{64.86}&66.96&51.53&60.58&62.58&61.24 & $p <$ 0.001\\
IRS (Ours) &  69.63&56.32&\textbf{60.43}&64.48&\textbf{69.06}&\textbf{54.37}&\textbf{61.13}&\textbf{64.24}&\textbf{62.17} & \textbf{Ref.}\\
\midrule
\rowcolor{gray!20} 
Joint & 75.32&63.03&62.08&63.28&68.76&52.67&64.80&63.37&64.32 & $p <$ 0.001 \\
\bottomrule
\end{tabular}
\end{adjustbox}
\label{tab:2-step_2}
\end{table}

\begin{figure}[t]
\centering 
\includegraphics[width=0.9\linewidth]{{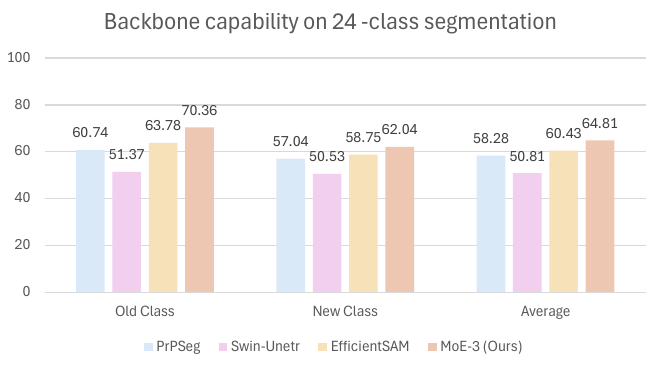}}
\caption{This figure shows the backbone capability on 24-class segmentation at step 2 for continual learning. The labels for all 24 classes are provided during the training. Dice similarity coefficient scores (\%) are reported.} 
\label{Fig.backbonecapability} 
\end{figure}

\begin{figure*}[t]
\centering 
\includegraphics[width=0.8\linewidth]{{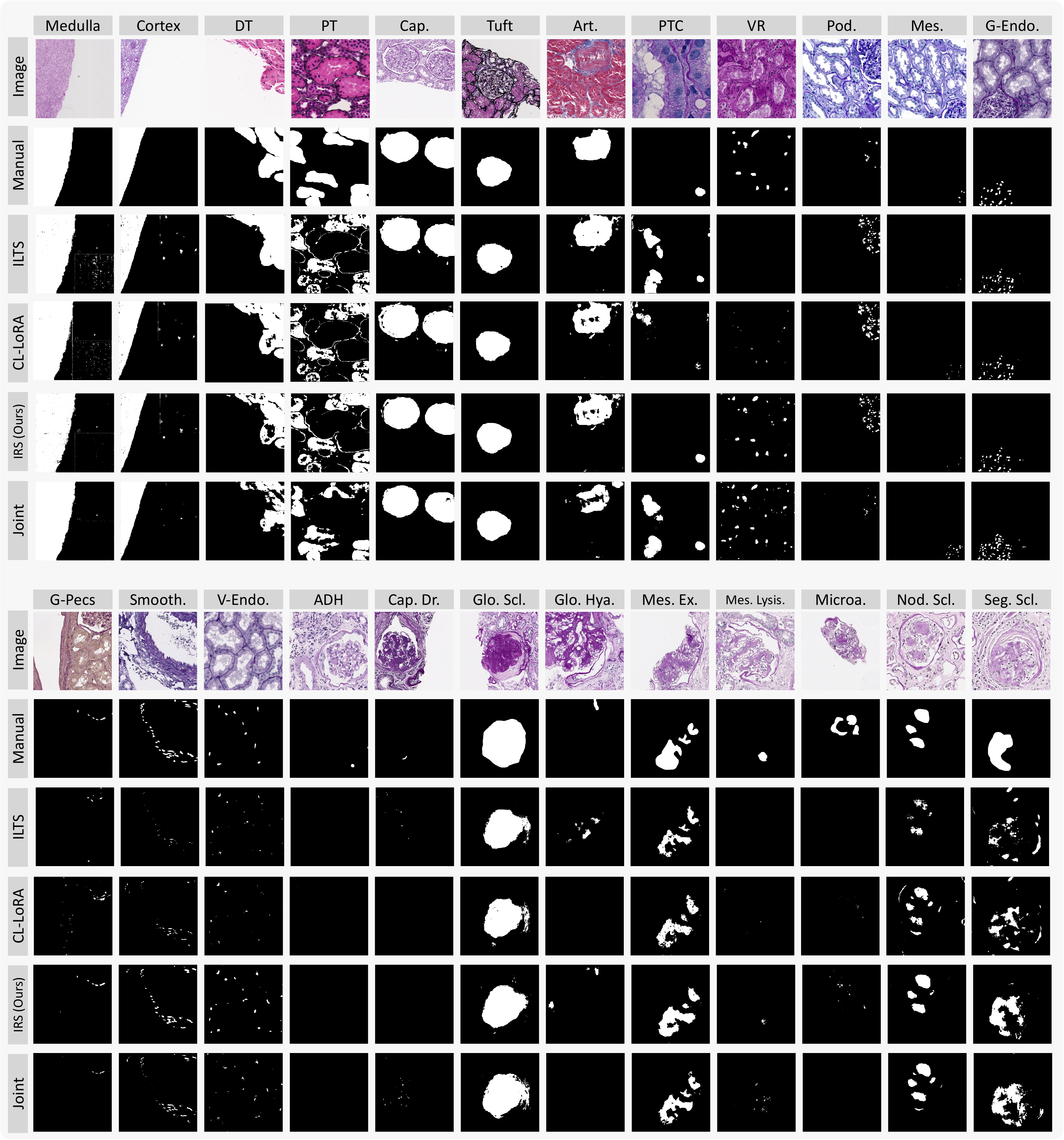}}
\caption{This figure presents the qualitative results of different continual learning approaches on all classes. The proposed method demonstrates superior knowledge distillation for old classes while achieving better learning on new classes.
}
\label{Fig.24classResults} 
\end{figure*}

\subsection{3-Step Continual Segmentation Performance}
\label{subsec:3step_results}

Fig.~\ref{Fig.Qualitative_3step} showcase the results from a 24-class continual segmentation evaluation in a 3-step setting, following the order of $D_1$-$D_2$-$D_3$. The results further demonstrate the capability and domain generalization of the proposed IRS in a more complex continual learning scenario, transitioning from coarse objects to fine objects and from the normal domain to the lesion domain.

\subsection{4-Step Continual Segmentation Performance}
\label{subsec:4step_results}

Table~\ref{tab:4-step} showcases the results from a class-by-class continual segmentation evaluation conducted in a 4-step setting. The results demonstrate the adaptability and flexibility of the proposed IRS across various data domain schemes.

\subsection{Ablation Study}
\label{subsec:ablationstudy}

\noindent\textbf{Knowledge Distillation Capacity:} Table~\ref{table:semidesign} indicates the capability of different semi-supervised losses for model knowledge distillation. The results show that the anatomy loss built from spatial-temporal relationship improves both old class memorization and new class learning, while the class tokens provide a bonus for the proposed prompt-driven backbone to achieve knowledge distillation.  

\noindent\textbf{Backbone Capability:} Table~\ref{table:backbonedesign}  in showcases the enhancements provided by our proposed prompt-driven dynamic MoE network for segmentation on $D_1$ in the first step. The results indicate that the dynamic tokens and the self-attention mechanism with the MoE dynamic head design boost the model's performance in segmenting objects at all levels. 

Fig.~\ref{Fig.backbonecapability} further showcases the performance improvements brought about by our proposed prompt-driven dynamic MoE network when utilizing all 24-class labels in the second step. Previous backbones proved insufficient as the number of classes increased. In contrast, the proposed backbone demonstrates superior segmentation performance for large-scale objects in the continual learning setting compared to individual backbones alone.

\noindent\textbf{Data Order:} To further evaluate the capability of the proposed method, we also report the continual learning segmentation results obtained using different data orders: a 2-step continual segmentation (\textbf{following the order of $D_2$ then $D_1$}) and a 3-step continual segmentation (\textbf{following the order of $D_2$-$D_1$-$D_3$}) are presented in Table~\ref{tab:2-step_2} and Fig.~\ref{Fig.Qualitative_3step}. 

\noindent\textbf{Model Complexity:} All methods were implemented with the same backbone, consisting of 19,194,168 parameters. The GPU memory usage for different methods is as follows:  
ILTS~\cite{michieli2019incremental}: 22,951 MiB;
CL-LoRA~\cite{chen2024low}: 16,950 MiB; 
IRS without anatomy: 22,674 MiB; 
IRS: 22,946 MiB.

\section{Conclusion}
\label{sec:conclusion}

In this paper, we introduced the IRS framework, addressing the challenges of class-incremental semantic segmentation in kidney pathology. We proposed a flexible and extensible incremental universal proposition matrix to model spatial-temporal relationships between old and new classes. This matrix guides the model to maintain and leverage knowledge of old classes when learning new ones. A prompt-driven dynamic MOE network architecture enhances the model's capability to handle an increased number of classes in continual learning, while minimizing architectural changes to ensure scalability and stability. By leveraging anatomical knowledge and dynamic network architectures, our method facilitates comprehensive and scalable segmentation across multiple biological scales and OOD domains.

\noindent\textbf{Acknowledgments.} This research was supported by the WCM Radiology AIMI Fellowship, NIH R01DK135597 (Huo), DoD HT9425-23-1-0003 (HCY), and the KPMP Glue Grant. This work was also supported by the Vanderbilt Seed Success Grant, the Vanderbilt Discovery Grant, and the VISE Seed Grant. This research was also supported by NIH grants R01EB033385, R01DK132338, REB017230, R01MH125931, and NSF 2040462. We extend gratitude to NVIDIA for their support by means of the NVIDIA hardware grant.

\bibliographystyle{IEEEtran}
\bibliography{tmi}

\end{document}